\documentclass[preprint,tightenlines,aps,floats,nofootinbib]{revtex4}  
\usepackage{epsf}

\newcommand{\be}{\begin{equation}}
\newcommand{\ee}{\end{equation}}
\newcommand{\bea}{\begin{eqnarray}}
\newcommand{\eea}{\end{eqnarray}}

\newcommand{\lsim}{
\mathrel{\hbox{\rlap{\hbox{\lower4pt\hbox{$\sim$}}}\hbox{$<$}}}}
\newcommand{\gsim}{
\mathrel{\hbox{\rlap{\hbox{\lower4pt\hbox{$\sim$}}}\hbox{$>$}}}}

\preprint{  
\hbox to \hsize{  
\hfill$\vcenter{\hbox{\bf MADPH-04-1400}
	        \hbox{\bf OKHEP-04-04}
	        \hbox{\bf UPR-1099T}
                \hbox{\bf hep-ph/0412136} 
                \hbox{December 2004}}$ } 
}  
  
\begin{document}  
  
\title{\vspace*{.75in}
Muon Anomalous Magnetic Moment\\in a Supersymmetric $U(1)'$ Model}

\author{
Vernon Barger$^1$, Chung Kao$^2$, Paul Langacker$^3$, and Hye-Sung Lee$^1$}
  
\affiliation{  
$^1$Department of Physics, University of Wisconsin,  
Madison, WI 53706 \\  
$^2$Department of Physics and Astronomy, University of Oklahoma,  
Norman,  OK 73019 \\
$^3$Department of Physics and Astronomy, University of Pennsylvania, 
Philadelphia, PA 19104
\vspace*{.5in}}
  
  
\thispagestyle{empty}  

\begin{abstract}  
 
We study the muon anomalous magnetic moment $a_\mu = (g_\mu - 2)/2$ in a supersymmetric $U(1)'$ model.
The neutralino sector has extra components from the superpartners of the $U(1)'$ gauge boson and the extra Higgs singlets that break the $U(1)'$ symmetry.
The theoretical maximum bound on the lightest neutralino mass is much smaller than that of the Minimal Supersymmetric Standard Model (MSSM) because of the mixing pattern of the extra components.
In a $U(1)'$ model where the $U(1)'$ symmetry is broken by a secluded sector (the $S$-model), $\tan\beta$ is required to be $\lsim 3$ to have realistic electroweak symmetry breaking.
These facts suggest that the $a_\mu$ prediction may be meaningfully different from that of the MSSM.
We evaluate and compare the muon anomalous magnetic moment in this model and the MSSM and discuss the constraints on $\tan\beta$ and relevant soft breaking terms.
There are regions of the parameter space that can explain the experimental deviation of $a_\mu$ from the Standard Model calculation and yield an acceptable cold dark matter relic density without conflict with collider experimental constraints.
\end{abstract}  

\maketitle

\newpage

\section{Introduction}
The anomalous magnetic moment of the muon $a_\mu = (g - 2)_\mu / 2$ is one of the most precisely measured physical quantities.
Its current value from the Brookhaven National Laboratory E821 experiment is \cite{BNLg-2, BNLg-2_develop}
\be
a_\mu(\mbox{exp}) = (11~659~208 \pm 6) \times 10^{-10},
\label{eq:a_exp}
\ee
which is a $2.4 \sigma$ deviation from the Standard Model (SM) prediction
\be
\Delta a_\mu \equiv a_\mu(\mbox{exp}) - a_\mu(\mbox{SM}) = (23.9 \pm 10.0) \times 10^{-10}
\label{eq:Delta_a}
\ee
when the hadronic vacuum polarization information is taken directly from the annihilation of $e^+ e^-$ to hadrons \cite{hadronic_from_ee} measured at CMD-2 \cite{CMD-2}.
The uncertainties involved in Eq. (\ref{eq:Delta_a}) are $7.2 \times 10^{-10}$ from the leading-order hadronic contribution \cite{hadLO}, $3.5 \times 10^{-10}$ from the hadronic light-by-light scattering \cite{lbl}, and $6 \times 10^{-10}$ from the $a_\mu$ experiment.
The indirect hadronic information from the hadronic $\tau$ decay gives a higher SM value that does not indicate a significant discrepancy with the SM (only a $0.9 \sigma$ deviation)\footnote{For a recent review of the various SM predictions and the $a_\mu$ discrepancies, see Ref. \cite{Passera}.}.
Recently released KLOE data \cite{KLOE} show an overall agreement with the CMD-2 data \cite{CMD-2}, confirming that there is a discrepancy between the hadronic contributions from the $e^+ e^-$ data and the $\tau$ data obtained from ALEPH, CLEO and OPAL \cite{Hocker}.

New physics is expected to exist at the TeV-scale to resolve various theoretical problems, including Higgs mass stabilization, and new physics could give a significant contribution to $a_\mu$ to explain the above deviation \cite{NewPhysics}.
There have been extensive studies of $a_\mu$ in supersymmetric (SUSY) models \cite{susy_amu}, which show that supersymmetry can naturally explain the deviation of Eq. (\ref{eq:Delta_a}).

The $a_\mu$ data constrains the SUSY parameters, including the sign of $\mu$ \cite{signmu} and upper limits on relevant scalar and fermion superpartner masses \cite{sleptonmass}.
In the minimal supergravity model (mSUGRA) or the Minimal Supersymmetric Standard Model (MSSM), the dominant additional contribution to $a_\mu$ comes from the first-order radiative corrections of the chargino-sneutrino and the neutralino-smuon loops;
it is
\be
\Delta a_\mu (\mbox{SUSY}) \sim 13 \times 10^{-10} \frac{\tan\beta~ \mbox{sign}(\mu)}{(M_{\mbox{\tiny SUSY}} / 100 \mbox{~GeV})^2}
\label{eq:Deltaamu}
\ee
in the limit that all the supersymmetric masses are degenerate at $M_{\mbox{\tiny SUSY}}$ \cite{moroi}.
The 2-loop corrections involve sfermion subloops or chargino/neutralino subloops and are at about the few percent level, although full calculations are not yet complete \cite{2loop}.
The discrepancy in Eq. (\ref{eq:Delta_a}) shows that a supersymmetry solution can be found if sign$(\mu) > 0$ and $M_{\mbox{\tiny SUSY}} \lsim 700$ GeV for $\tan\beta \lsim 50$, in the limit that supersymmetric masses are degenerate.
The deviation of $a_\mu$ similarly gives constraints on the parameters of other new physics models including the mass of a second generation leptoquark \cite{Cheung:prd.64.033001}, the mass of the heavy photon in the little Higgs model \cite{Park:hep-ph/0306112} and the compactification scale of an extra dimension \cite{extraDimension}.

Given that $a_\mu$ has been a powerful tool for constraining the new physics models, due to the accuracy of its measurement and the SM evaluation, it is interesting to pursue what $a_\mu$ can tell about recently emerging models.
The recent idea of split supersymmetry assumes large masses (e.g., $10^{10}$ GeV) for scalar superpartners (sleptons, squarks) while keeping fermionic superpartners (gauginos, higgsinos) at the TeV-scale \cite{splitSUSY}.
The large masses of the smuon and sneutrino would make the chargino-sneutrino and neutralino-smuon loop contributions to $a_\mu$ negligible; the split supersymmetry model would be rejected if the deviation of $a_\mu$ is in fact real.

Another interesting TeV-scale new physics model is the supersymmetric $U(1)'$ model \cite{onesinglet, smodel}.
It has a structure similar to the MSSM but has an extra $U(1)$ gauge symmetry ($U(1)'$), which is spontaneously broken at the TeV-scale by one or multiple Higgs singlets.
This model can provide natural solutions to some of the difficulties the MSSM faces, including the explanation of the electroweak scale of the $\mu$ parameter ($\mu$-problem \cite{muproblem}) and the lack\footnote{The required strong first-order phase transition for EWBG is allowed in the MSSM only if the light Higgs mass is only slightly above the LEP experimental bound and the light stop mass is smaller than the top mass \cite{EWBG_MSSM}.} of a sufficiently strong first-order phase transition for electroweak baryogenesis (EWBG) \cite{EWBG_U1}.
The Next-to-Minimal Supersymmetric Standard Model (NMSSM) \cite{NMSSM} can also resolve the $\mu$-problem but its discrete $\bf Z_3$ symmetry invokes a cosmological domain wall problem \cite{domainwall}; a variant which avoids this problem is discussed in Ref. \cite{nMSSM, Wagner}.

Besides the bottom-up reasons to introduce an additional $U(1)$ symmetry to supplement the MSSM, many new physics models, including grand unified theories (GUTs), extra dimensions \cite{extradim}, superstrings \cite{superstring}, little Higgs \cite{littleHiggs}, dynamical symmetry breaking \cite{dynamical} and Stueckelberg mechanism models \cite{Stueckelberg} predict extra $U(1)$ symmetries or gauge bosons.
The newly introduced particles such as the $U(1)'$ gauge boson ($Z'$) and the $U(1)'$ breaking Higgs singlet ($S$) and their superpartners $Z'$-ino ($\tilde Z'$) and singlino ($\tilde S$), alter the Higgs and neutralino spectra.
The modified Higgs spectrum \cite{Higgs_U1} and the neutralino relic density \cite{relic_U1} have been recently studied in the $U(1)'$ model with a secluded $U(1)'$ symmetry breaking sector \cite{smodel}, and the difference in the predictions from the MSSM detailed.

There have been studies of the muon anomalous magnetic moment in models with additional gauge groups or $E_6$ GUT \cite{Leveille, E6_g-2, exotic_g-2}.
They mostly concentrated on the loops including the $Z'$ or the exotic quarks in the model and their superpartners, and constraints were obtained on their masses or couplings.
To explain the $a_\mu$ deviation, the $Z'$ masses should be typically smaller than the experimental limits of $M_{Z'} \gsim 500 \sim 800$ GeV from direct searches at the Tevatron \cite{CDF, PDG}.

In this paper we quantitatively study a supersymmetric $U(1)'$ model with a secluded $U(1)'$ symmetry breaking sector (the $S$-model) \cite{smodel} to see how the extended neutralino sector contribution to $a_\mu$ is different from the MSSM prediction.
The superpotential in this model is
\be
W = h_s S H_1 H_2 + \lambda_s S_1 S_2 S_3,
\ee
where $h_s$ and $\lambda_s$ are dimensionless parameters.
The $\mu$-problem is solved by replacing the $\mu$ term by an effective $\mu$ parameter
\be
\mu_{\rm eff} = h_s s / \sqrt{2}
\ee
where $s / \sqrt{2}$ is the vacuum expectation value (VEV) of the Higgs singlet $S$ that acquires a VEV at the electroweak or TeV scale.
The $Z'$ has a large mass generated by the Higgs singlet fields $S_{1,2,3}$, which acquire large (TeV scale) VEVs for small $\lambda_s$ because of an almost $F$ and $D$ flat direction.
The extra Higgs singlets allow $\mu_{\rm eff}$ to be at the electroweak scale while keeping the $Z'$ heavier than the experimental limit.

The electroweak symmetry breaking is driven by electroweak scale trilinear soft terms, leading to small values for $\tan\beta \equiv v_2 / v_1$ ($\tan\beta \sim 1$ to $3$), while solutions without unwanted global minima at $\left< H_i^0 \right> = 0$ typically have $\left< S \right> \lsim 1.5 \left< H_i^0 \right>$ \cite{smodel}.
A small $\tan\beta$ implies a problem in the MSSM; for example, the light Higgs mass needs large $\tan\beta$ (for reasonable superpartner masses) to satisfy the LEP bound of $m_h > 115$ GeV.
The mixing of the Higgs doublets and singlets in the $S$-model, however, lowers the LEP $m_h$ bound significantly \cite{Higgs_U1}.
Furthermore, the maximum theoretical value for $m_h$ for a given $\tan\beta$ is increased, both by $F$-terms (similar to effects in the NMSSM \cite{NMSSM, Higgs_U1}) and by $D$-terms \cite{Dterms}.
As a result, $\tan\beta \sim 1$ to $3$ is experimentally allowed in the $S$-model.

The neutralinos in the $S$-model have extra components of the $Z'$-ino and singlinos.
The lightest neutralino of the $U(1)'$ model is often very light because of the mixing pattern of the extra components.
Eq. (\ref{eq:Deltaamu}) suggests that $a_\mu$ in this model may be significantly different from that of the MSSM. 
We numerically investigate the differences and obtain constraints on $\tan\beta$ and the relevant soft breaking terms.
We do not include $Z'$ loops since the $Z'$ mass bounds from direct searches at colliders imply negligible effects.

In Section \ref{section:formalism} we describe the formalism to compute the chargino and neutralino contributions in the $S$-model.
In Section \ref{section:analysis}, we give the numerical analysis of $a_\mu$ and make comparisons to the MSSM results, before the conclusion in Section \ref{section:conclusion}.

\section{Supersymmetric Contributions to $a_\mu$ with $U(1)'$ symmetry}
\label{section:formalism}
In this section we describe the neutralino and chargino contributions to $a_\mu$ in the $S$-model.
The formalism is similar to  the MSSM with a straightforward extension.
We assume that the VEVs of the $S_{1,2,3}$ are large compared to the VEVs of other Higgs fields ($H_1^0$, $H_2^0$ and $S$) and that their singlino components essentially decouple \cite{smodel, Higgs_U1}.

\subsection{Neutralino Contribution}
Ignoring the $\tilde S_{1,2,3}$ singlinos, the neutralino mass matrix in the basis $\left\{\tilde{B}\right.$, $\tilde{W}_3$, $\tilde{H}_1^0$, $\tilde{H}_2^0$, $\tilde{S}$, $\left.\tilde{Z'} \right\}$ is given by
\bea
M_{\chi^0}= 
\left( \matrix{M_1 & 0 & - g_1 v_1 / 2 & g_1 v_2 / 2 & 0 & 0 \cr
0 & M_2 & g_2 v_1 / 2 & - g_2 v_2 / 2 & 0 & 0 \cr
- g_1 v_1 / 2 & g_2 v_1 / 2 & 0 & - h_s s / \sqrt{2} & - h_s v_2 / \sqrt{2} & g_{Z'} Q'(H_1^0) v_1 \cr
g_1 v_2 / 2 & - g_2 v_2 / 2 & - h_s s / \sqrt{2} & 0 & - h_s v_1 / \sqrt{2} & g_{Z'} Q'(H_2^0) v_2 \cr
0 & 0 & - h_s v_2 / \sqrt{2} & - h_s v_1 / \sqrt{2} & 0 & g_{Z'} Q'(S) s \cr
0 & 0 & g_{Z'} Q'(H_1^0) v_1 & g_{Z'} Q'(H_2^0) v_2 & g_{Z'} Q'(S) s & M_{1'}} \right)
\label{eqn:neutralinomassmatrix}
\eea
where $e = g_1\cos\theta_W = g_2\sin\theta_W$;
$g_{Z'}$ is the $U(1)'$ gauge coupling constant, for which we take the GUT motivated value of $g_{Z'} = \sqrt{5/3} g_1$.
$Q'$ is the $U(1)'$ charge, and the anomaly-free charge assignments based on $E_6$ GUT can be found in Ref. \cite{E6model}.

The VEVs of the Higgs doublets are $\langle H_i^0 \rangle \equiv 
\frac{v_i}{\sqrt{2}}$ with $\sqrt{v_1^2 + v_2^2} \simeq 246$ GeV.
The diagonalization of the mass matrix can be accomplished using a unitary matrix $N$,
\be
N^T M_{\chi^0} N = Diag(M_{\chi^0_1}, M_{\chi^0_2}, M_{\chi^0_3}, M_{\chi^0_4}, M_{\chi^0_5}, M_{\chi^0_6}).
\ee
The first $4\times4$ ($5\times5$) submatrix of Eq. (\ref{eqn:neutralinomassmatrix}) is the MSSM (NMSSM) limit.
Due to the singlino addition, there exists a kind of see-saw mechanism that makes the lightest neutralino very light \cite{lightneutralino, Wagner}, less than $100$ GeV in the case of $M_{1'} \gg M_1$ (where the $\tilde{Z'}$ practically decouples and the mass matrix becomes the NMSSM limit) \cite{relic_U1}.
We will consider both this limit and that in which the gaugino mass unification relation, $M_{1'} = M_1 = \frac{5}{3} \frac{g_1^2}{g_2^2} M_2 \simeq 0.5 M_2$, is satisfied.

\begin{figure}[t]
\begin{minipage}[b]{1\textwidth}
\centering\leavevmode
\epsfbox{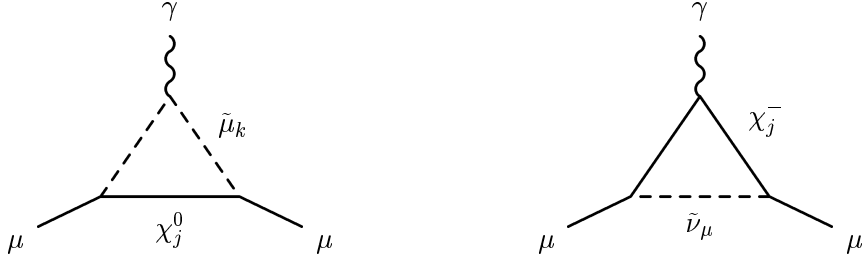}
\end{minipage}
\label{fig:diagram}
\caption{Supersymmetric contributions to $a_\mu$ involving charginos and neutralinos}
\end{figure}

The smuon mass-squared matrix is given by
\bea
M_{\tilde\mu}^2 =
\left( \matrix{M_{LL}^2 & M_{LR}^2 \cr
M_{RL}^2 & M_{RR}^2} \right).
\label{eq:smuonmass}
\eea
\bea
M_{LL}^2 &=& m_{\tilde L}^2 + m_\mu^2 + (T_{3 \mu} - Q_\mu \sin^2\theta_W) M_Z^2 \cos2\beta \\
M_{RR}^2 &=& m_{\tilde E}^2 + m_\mu^2 + (Q_\mu \sin^2\theta_W) M_Z^2 \cos2\beta \\
M_{LR}^2 &=& m_\mu (A_\mu^* - \mu_{\rm eff} \tan\beta) \label{eq:MLR2} \\
M_{RL}^2 &=& (M_{LR}^2)^* \label{eq:MRL2}
\eea
Its diagonalization can be accomplished through the unitary matrix $D$ as
\be
D^\dagger M_{\tilde\mu}^2 D = Diag(M_{\tilde\mu_1}^2,M_{\tilde\mu_2}^2).
\ee

The LEP2 SUSY Working Group analysis found $m_{\tilde \mu_R} \gsim 95$ GeV from $\tilde \mu \to \mu \chi^0_{1,2}$ searches \cite{LEPSUSY}.
Since $m_\mu$ is small compared to supersymmetric parameters, the off-diagonal terms of Eq. (\ref{eq:MLR2}) and Eq. (\ref{eq:MRL2}) are small and hence the mixing is small.
For $\tan\beta \gsim 1$, both $m_{\tilde L}^2$, $m_{\tilde E}^2 \gsim (95~ \mbox{GeV})^2$ are required to give $m_{\tilde \mu_1} \gsim 95$ GeV from the tree-level mass matrix of Eq. (\ref{eq:smuonmass}), while for $\tan\beta \lsim 1$, somewhat larger values of these parameters are required.
We require both $m_{\tilde L}^2$ and $m_{\tilde E}^2$ to be larger than $(100~ \mbox{GeV})^2$ and take $A_\mu = 0$ in our numerical analysis.

The neutralino contribution to $a_\mu$ is then \cite{susy_amu}
\be
a_\mu(\chi^0) = a_\mu^1(\chi^0) + a_\mu^2(\chi^0)
\ee
where
\bea
a_\mu^1(\chi^0) &=& \sum_{j=1}^6 \sum_{k=1}^{2} \frac{m_\mu}{8 \pi^2 M_{\chi_j^0}} Re[L_{jk} R_{jk}^*] F_1(\frac{M_{\tilde{\mu}_k}^2}{M_{\chi_j^0}^2}) \\
a_\mu^2(\chi^0) &=& \sum_{j=1}^{6} \sum_{k=1}^{2} \frac{m_\mu^2}{16 \pi^2 M_{\chi_j^0}^2} \left( |L_{jk}|^2 + |R_{jk}|^2 \right) F_2(\frac{M_{\tilde{\mu}_k}^2}{M_{\chi_j^0}^2})
\eea
with the following $\mu$-$\tilde\mu$-$\chi^0$ chiral coupling:
\bea
L_{j k} &=& \frac{1}{\sqrt{2}} \left( g_1 Y_{\mu_L} N_{1 j}^* - g_2 N_{2 j}^* + g_{Z'} Q'(\mu_L) N_{6 j}^* \right) D_{1 k} + \frac{\sqrt{2} m_\mu}{v_1} N_{3 j}^* D_{2 k} \\
R_{j k} &=& \frac{1}{\sqrt{2}} \left( g_1 Y_{\mu_R} N_{1 j} + g_{Z'} Q'(\mu_R) N_{6 j} \right) D_{2 k} + \frac{\sqrt{2} m_\mu}{v_1} N_{3 j} D_{1 k}.
\eea
$Y_{\mu_L} = -1$, $Y_{\mu_R} = 2$ are hypercharges and the terms with $g_{Z'}$ coupling are the additional contributions from the $U(1)'$.
Since $m_\mu$ is small compared to the  supersymmetric masses, we can approximate it as zero in the loop integral functions to obtain
\bea
F_1 (x) &=& \frac{1}{2} \frac{1}{(x-1)^3} (1 - x^2 + 2x\ln x) \\
F_2 (x) &=& \frac{1}{6} \frac{1}{(x-1)^4} (-x^3 + 6x^2 - 3x - 2 - 6x\ln x).
\eea

\subsection{Chargino Contribution}
The chargino mass matrix is given by
\bea
M_{\chi^\pm} =
\left( \matrix{M_2 & \sqrt{2} M_W \sin\beta \cr
\sqrt{2} M_W \cos\beta & h_s s / \sqrt{2}} \right).
\label{eq:charginomass}
\eea
It is essentially the same as in the MSSM except that $\mu$ is replaced by $\mu_{\rm eff} = h_s s / \sqrt{2}$.
$M_{\chi^\pm}$ can be diagonalized by two unitary matrices $U$ and $V$ as
\be
U^* M_{\chi^\pm} V^{-1} = Diag(M_{\chi^\pm_1},M_{\chi^\pm_2}).
\ee
The LEP light chargino mass limit $M_{\chi^-} \gsim 104$ GeV \cite{charginomass} gives constraints on $M_2$ and $\mu_{\rm eff}$ for a fixed value of $\tan\beta$.
The sneutrino mass-squared is given by
\be
M_{\tilde \nu_\mu}^2 = m_{\tilde L}^2 + T_{3 \nu} M_Z^2 \cos 2\beta.
\ee
As in the MSSM calculations, we do not include the right-handed neutrino or its superpartner.

The chargino loop contribution to $a_\mu$ is \cite{susy_amu}
\be
a_\mu(\chi^-) = a_\mu^1(\chi^-) + a_\mu^2(\chi^-)
\label{eq:chargino_amu}
\ee
where
\bea
a_\mu^1(\chi^-) &=& \sum_{j=1}^{2} \sum_{k=1}^{1} \frac{m_\mu}{8 \pi^2 M_{\chi_j^-}} Re[L_{jk} R_{jk}^*] F_3(\frac{M_{\tilde\nu_\mu}^2}{M_{\chi_j^-}^2}) \\
a_\mu^2(\chi^-) &=& - \sum_{j=1}^{2} \sum_{k=1}^{1} \frac{m_\mu^2}{16 \pi^2 M_{\chi_j^-}^2} \left( |L_{jk}|^2 + |R_{jk}|^2 \right) F_4(\frac{M_{\tilde\nu_\mu}^2}{M_{\chi_j^-}^2})
\eea
with the chiral $\mu$-$\tilde\nu_\mu$-$\chi^-$ couplings
\bea
L_{j1} = \frac{\sqrt{2} m_\mu}{v_1} U_{j2}^*, \qquad R_{j1} = - g_2 V_{j1}.
\eea
The loop integral functions are
\bea
F_3 (x) &=& -\frac{1}{2} \frac{1}{(x-1)^3} (3x^2 - 4x + 1 - 2x^2\ln x) \\
F_4 (x) &=& -\frac{1}{6} \frac{1}{(x-1)^4} (2x^3 + 3x^2 - 6x + 1 -6x^2\ln x).
\eea

\section{Analysis}
\label{section:analysis}
\begin{figure}[t]
\begin{minipage}[c]{\textwidth}
\centering\leavevmode
\epsfxsize=3.2in
\epsfbox{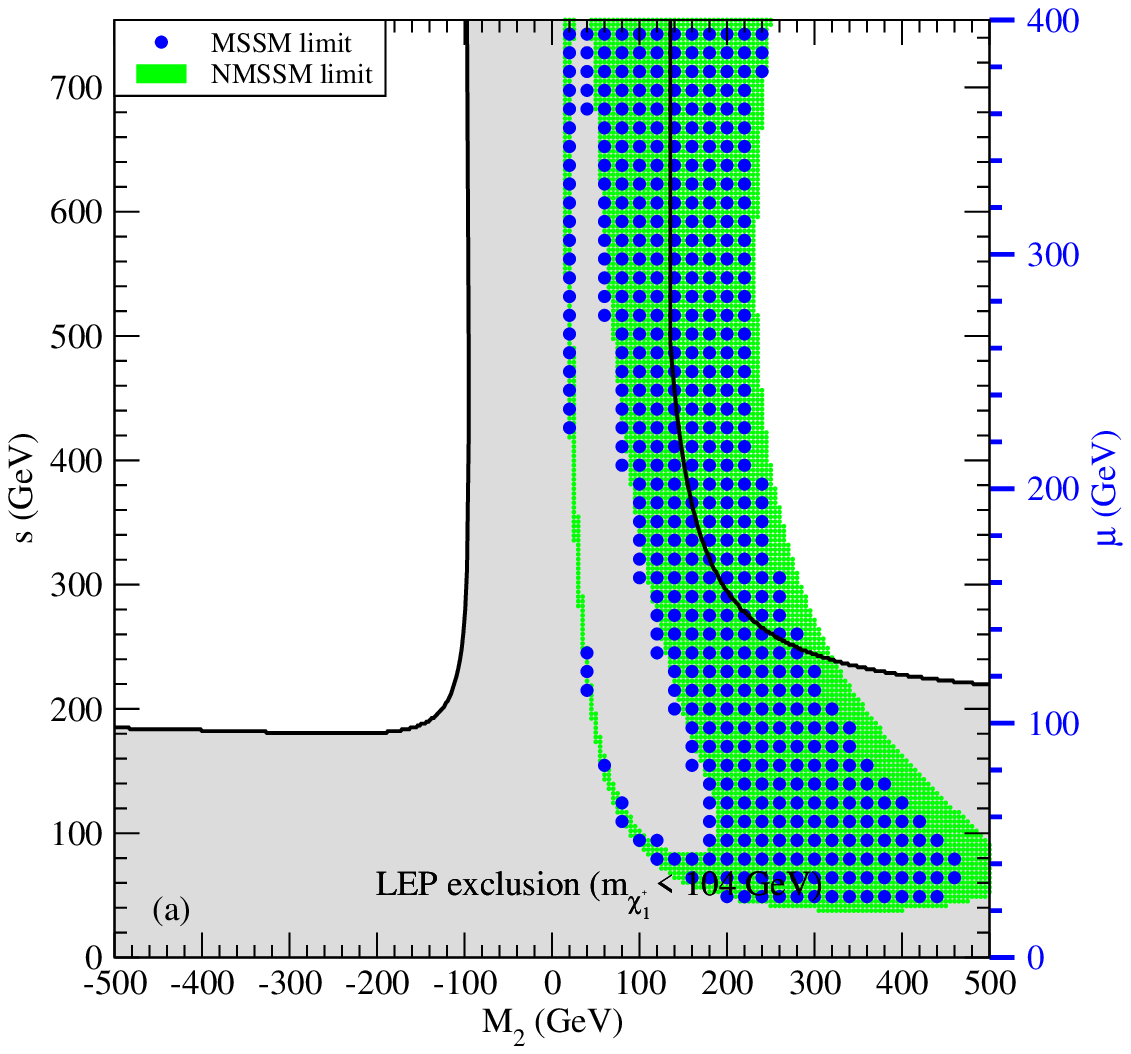}
\epsfxsize=3.2in
\epsfbox{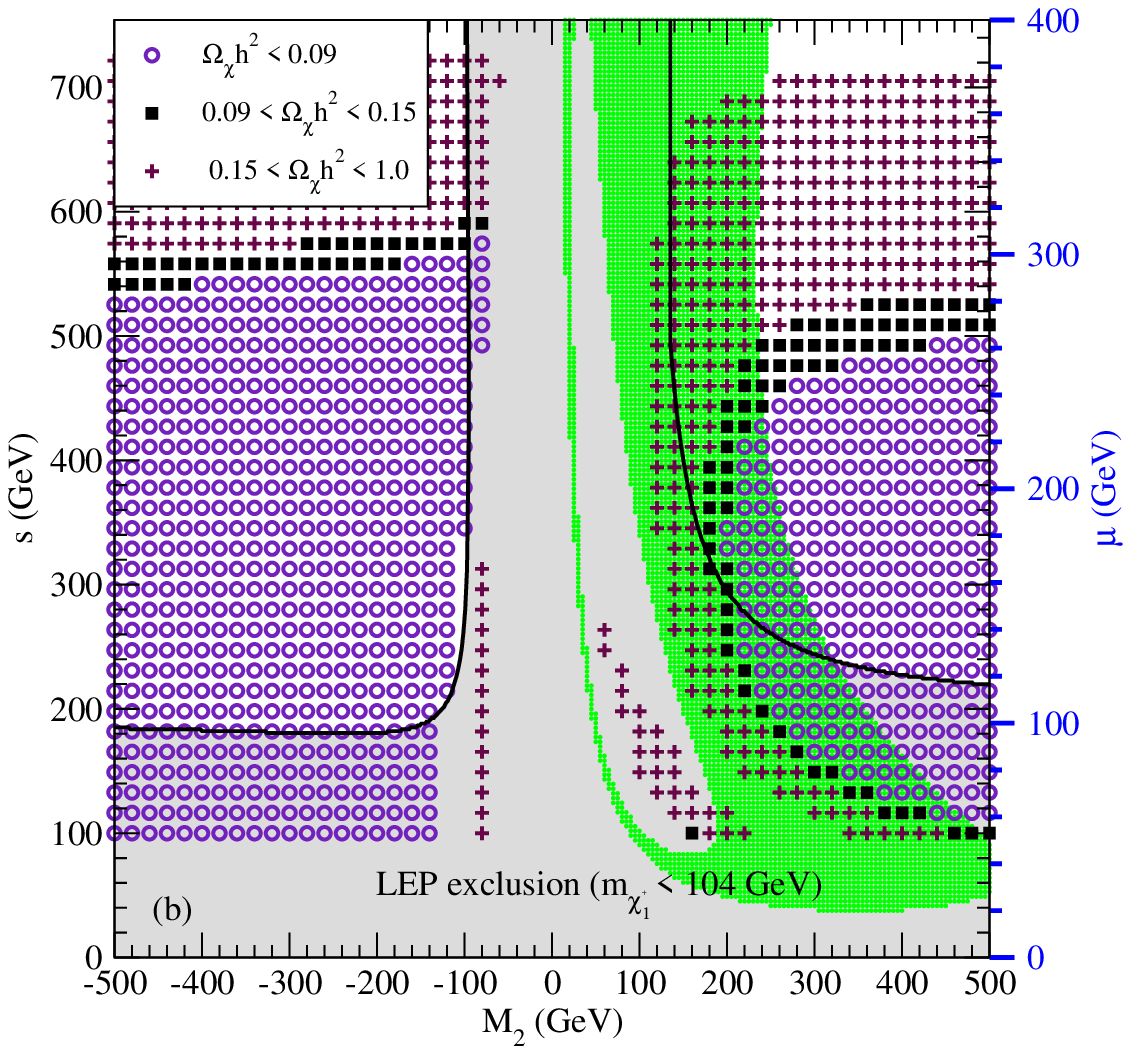}
\end{minipage}
\caption{(a) $\Delta a_\mu$ in the MSSM [filled circles] and the NMSSM limit ($M_{1'} \gg M_2$) of the $S$-model [dark shading] for $\tan\beta = 2.5$ and $m_{\rm smuon} = 100~ \mbox{GeV}$. For the $S$-model, $h_s = 0.75$ and $\eta$-model charges are assumed. The region outside the NMSSM region has $\Delta a_\mu \times 10^{10} < 13.9$, while the island inside the NMSSM region (around $M_2 \sim 100$ GeV) has $\Delta a_\mu \times 10^{10} > 33.9$. The lightly shaded area outlined by solid curves is excluded by the LEP chargino mass limit $M_{\chi^-} > 104$ GeV. The $\Delta a_\mu$ have values $13.9 \le \Delta a_\mu \times 10^{10} \le 33.9$ favored by Eq. (\ref{eq:Delta_a}). The models give similar $a_\mu$ results and all the models allow small $\tan\beta$ for $m_{\rm smuon}$ small ($\sim 100$ GeV). (b) The solution space for the acceptable neutralino relic density \cite{relic_U1} and the $(g-2)_\mu$ deviation are shown together for the same NMSSM limit. The filled squares have the WMAP $3\sigma$ allowed range of $0.09 < \Omega_{\chi^0} h^2 < 0.15$, the open circles have $\Omega_{\chi^0} h^2 < 0.09$ and the crosses have $0.15 < \Omega_{\chi^0} h^2 < 1.0$.}
\label{fig:contour_M2_mu}
\end{figure}

The MSSM can explain the $2.4 \sigma$ deviation between the E821 experiment and the SM prediction for most values of $\tan\beta$.
In this section, we compare predictions from the MSSM with those of the $S$-model.

\begin{figure}[t]
\begin{minipage}[c]{\textwidth}
\begin{minipage}[b]{.49\textwidth}
\centering\leavevmode
\epsfxsize=3in
\epsfbox{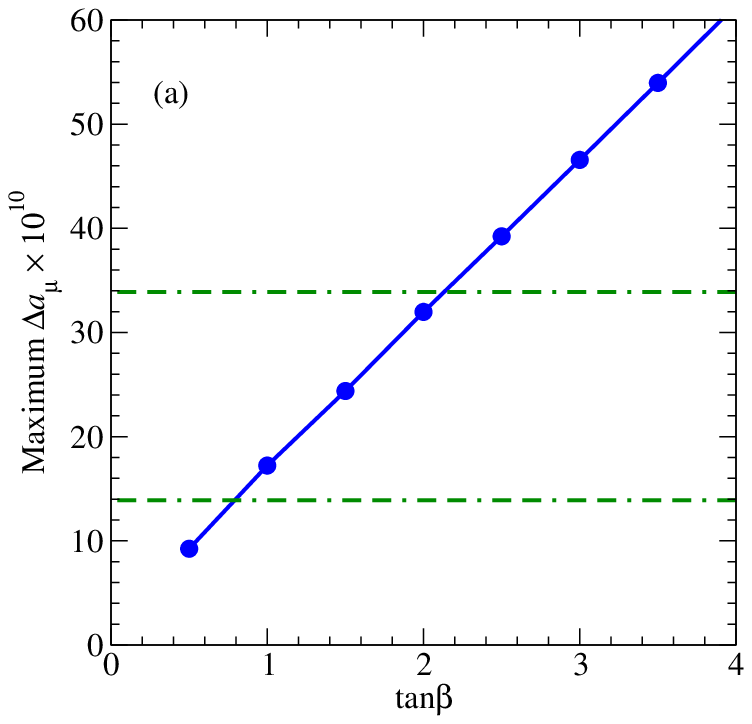}
\end{minipage}
\hfill
\begin{minipage}[b]{.49\textwidth}
\centering\leavevmode
\epsfxsize=3in
\epsfbox{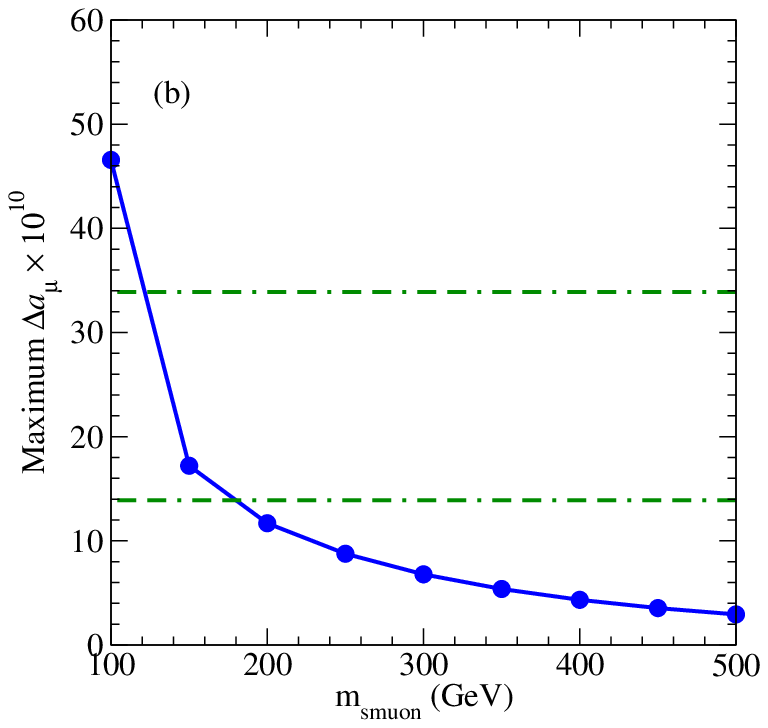}
\end{minipage}
\end{minipage}
\caption{Maximum $\Delta a_\mu$ versus (a) $\tan\beta$ and (b) $m_{\rm smuon}$ in the $S$-model with $M_{1'} = M_1$.
The MSSM result and the NMSSM limit $(M_{1'} \gg M_2)$ are nearly indistinguishable from the $M_{1'} = M_1$ curve and are not plotted.
$M_2$ is scanned from $100$ GeV to $1000$ GeV, and $\mu$($\mu_{\rm eff}$) from $100$ GeV to $1000$ GeV. $m_{\rm smuon}$ is scanned from $100$ GeV to $500$ GeV for (a), and $\tan\beta$ from $1$ to $3$ for (b).
Horizontal dashed-dot lines are boundaries of the measured deviation $\Delta a_\mu \times 10^{10} = 13.9 \sim 33.9$ of Eq. (\ref{eq:Delta_a}).}
\label{fig:amu_tanbMsoft}
\end{figure}

\begin{figure}[t]
\begin{minipage}[c]{\textwidth}
\begin{minipage}[b]{.49\textwidth}
\centering\leavevmode
\epsfxsize=3in
\epsfbox{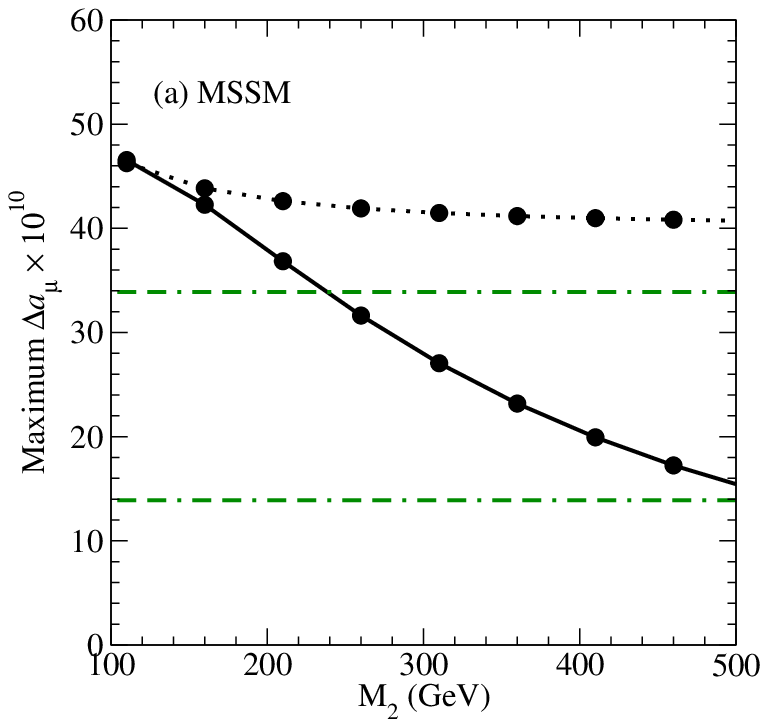}
\end{minipage}
\hfill
\begin{minipage}[b]{.49\textwidth}
\centering\leavevmode
\epsfxsize=3in
\epsfbox{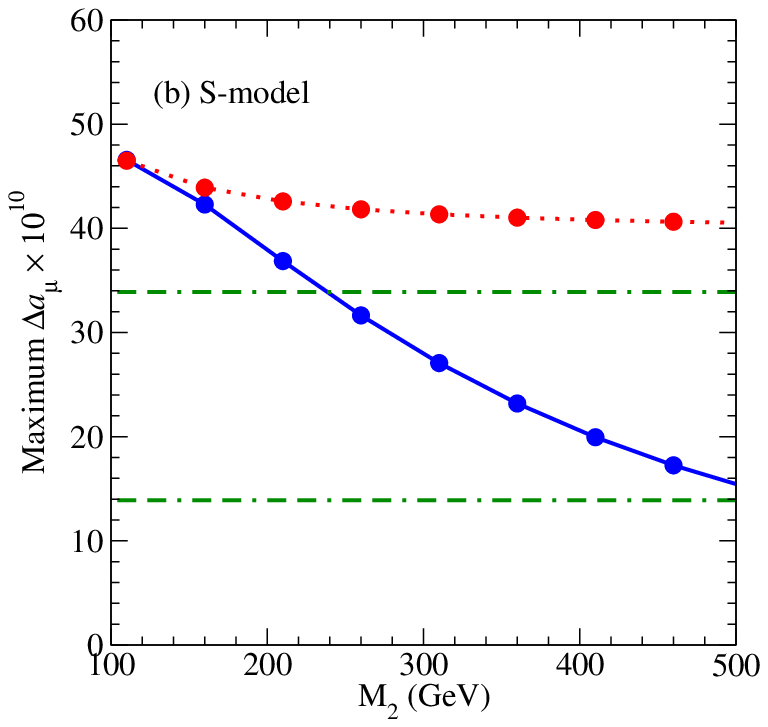}
\end{minipage}
\end{minipage}
\caption{Maximum $\Delta a_\mu$ versus $M_2$ in (a) the MSSM and (b) the $S$-model. The dotted curves take $M_1$ as a free parameter while the solid ones follow the gaugino mass unification relation ($M_{1'} = M_1 \simeq 0.5 M_2$). $\mu$($\mu_{\rm eff}$) is scanned from $100$ GeV to $1000$ GeV, $M_1$ (for dotted curves) from $50$ GeV to $1000$ GeV, $\tan\beta$ from $1$ to $3$, and $m_{\rm smuon}$ from $100$ GeV to $500$ GeV. When $M_1$ is a free parameter, we take $M_{1'} \gg M_2$ (NMSSM limit) for the $S$-model (dotted curve).
In both models, the maximum values of $\Delta a_\mu$ occur for $M_1 \sim 50$ GeV when $M_1$ is taken as a free parameter.}
\label{fig:amu_M2}
\end{figure}

Figure \ref{fig:contour_M2_mu} (a) shows $\Delta a_\mu$ in the $M_2$-$s$ (also $M_2$-$\mu$) plane for the MSSM [filled circles], the $S$-model and its NMSSM limit [dark shading], with $\tan\beta = 2.5$ and $m_{\rm smuon} \equiv m_{\tilde L} = m_{\tilde E} = 100$ GeV.
The comparison is made for $\mu = \mu_{\rm eff}$.
The corresponding plot for the $S$-model with $M_{1'} = M_1$ is hardly distinguishable from that of the NMSSM limit so it is not shown in the figure.
The plots are the parameter spaces that can give values $\Delta a_\mu \times 10^{10} = 13.9 \sim 33.9$ favored by Eq. (\ref{eq:Delta_a}).
The parameter space excluded by the LEP chargino mass limit of $M_{\chi^-} > 104$ GeV is shown as the lightly shaded area outlined by solid curves.
Throughout the following analysis (Figures \ref{fig:amu_tanbMsoft} and \ref{fig:amu_M2}), we do not include the parameter points that violate the LEP chargino mass constraint.
This figure shows that both the MSSM\footnote{It should be emphasized that the MSSM cannot accept this small $\tan\beta$ if we consider other constraints: the light Higgs mass would be too small to be compatible with the LEP Higgs mass bound of $m_h > 115$ GeV.} and the $S$-model (including the NMSSM limit) can explain the $\Delta a_\mu$ data with small $\tan\beta$ while satisfying $m_{\rm smuon} \gsim 100$ GeV from the experimental bound on the scalar muon mass.
The $S$-model gives only a slightly larger area in the parameter space than the MSSM. 
For the $S$-model, $h_s = 0.75$, and the $E_6$ motivated $\eta$-model\footnote{The $\eta$-model is a $U(1)'$ model that is produced with a unique set of charge assignments when $E_6$ is broken to a rank-5 group.}
charge assignments are assumed \cite{E6model}.

It is interesting to note that in the $M_{1'} \gg M_2$ case\footnote{The CDM expectations for smaller $M_{1'}$ have not yet been examined.} (NMSSM limit), a sizable part of the $a_\mu$ $2.4 \sigma$ deviation solution area overlaps the solution area which reproduces the observed cold dark matter (CDM) relic density in the same framework and limit \cite{relic_U1} as shown in Figure \ref{fig:contour_M2_mu} (b).
The filled squares have the WMAP $3\sigma$ allowed range of $0.09 < \Omega_{\chi^0} h^2 < 0.15$, the open circles have $\Omega_{\chi^0} h^2 < 0.09$ and the crosses have $0.15 < \Omega_{\chi^0} h^2 < 1.0$.
In the event that the $\Delta a_\mu$ deviation from the SM is real, this CDM result enhances the viability of the model.
The chargino mass allowed by both the $a_\mu$ deviation and the relic density is $104$ GeV $\lsim M_{\chi^\pm} \lsim 220$ GeV, where the lower bound is the present experimental limit.
Since the chargino mass could be only slightly larger than the present LEP limit, a search for the SUSY trilepton signal at the Tevatron Run II will be very interesting \cite{trilepton}.
It should be noted that only the $Z$-pole annihilation channel was considered to show this model could reproduce the acceptable relic density.
There could be a larger solution space when more channels are considered.

We now consider the limits on $\tan\beta$ and $m_{\rm smuon}$ that allow the favored range of $\Delta a_\mu$.
Figure \ref{fig:amu_tanbMsoft} (a) shows the maximum $\Delta a_\mu$ as a function of $\tan\beta$ in the $S$-model (with $M_{1'} = M_1$).
The MSSM curve is nearly the same, as is the NMSSM limit ($M_{1'} \gg M_2$) of the $S$-model.
$M_2$ is scanned from $100$ GeV to $1000$ GeV, $\mu$($\mu_{\rm eff}$) from $100$ GeV to $1000$ GeV (negative $\mu$ is nearly irrelevant for producing positive $\Delta a_\mu$), and $m_{\rm smuon}$ from $100$ GeV to $500$ GeV.
This figure demonstrates the almost linear dependence on $\tan\beta$, as in Eq. (\ref{eq:Deltaamu}), and shows that even with small $\tan\beta$, both models are able to produce the favored values of $\Delta a_\mu$.
Smaller $\tan\beta$, however, has less supersymmetric parameter space that explains the deviation.
The maximum values of $\Delta a_\mu$ have $m_{\rm smuon} \sim 100$ GeV (the lowest scan value).

Since the chargino contribution is the same in the MSSM and the $S$-model, the dominant difference in $a_\mu$ would come from the lightest neutralino ($\chi^0_1$) contribution.
The mass and the coupling of $\chi^0_1$ in the $S$-model is similar to that of the MSSM.
In the NMSSM limit ($M_{1'} \gg M_1$), $\chi^0_1$ is often significantly lighter than that of the MSSM but it is mostly singlino-like which couples to the muon only through mixing; $\chi^0_2$ is mostly similar to $\chi^0_1$ of the MSSM.

Figure \ref{fig:amu_tanbMsoft} (b) shows the maximum $\Delta a_\mu$ as a function of $m_{\rm smuon}$ in the $S$-model; the MSSM curve is nearly identical when $\tan\beta$ is scanned only from $1$ to $3$.
The maximum points have $\tan\beta \sim 3$ (the highest scan value).
A large value of $\tan\beta$, as more general $U(1)'$ models allow, would always ensure a sufficiently large maximum $\Delta a_\mu$ for a given $m_{\rm smuon}$.
This figure demonstrates the roughly inverse-squared dependence on $m_{\rm smuon}$ as in Eq. (\ref{eq:Deltaamu}).
It also shows that, for small $\tan\beta$, the scalar muon should be light ($100$ GeV $\lsim M_{\tilde \mu} \lsim 180$ GeV) to explain the $a_\mu$ deviation.
This could in principle lead to a concern for the neutralino cold dark matter candidate, since the scalar muon could be light enough to be the lightest supersymmetric particle (LSP).
A charged LSP would conflict with the observational absence of exotic isotopes.
In the $S$-model, however, the lightest neutralino is usually very light, e.g., $M_{\chi^0_1} \lsim 100$ GeV, while it produces the acceptable range of the relic density for a wide range of the parameter space \cite{relic_U1}.
One can therefore have a light slepton with an even lighter neutralino as the LSP.
A light slepton will be observable in future accelerators such as the CERN Large Hadron Collider (LHC) and International Linear Collider (ILC).
Light sleptons could be detected easily at a linear collider of a moderate energy of $500$ GeV.

Figure \ref{fig:amu_M2} shows the maximum $\Delta a_\mu$ as a function of $M_2$ in both models.
For the dotted curves, we do not impose the gaugino mass unification assumption of $M_{1'} = M_1 \simeq 0.5 M_2$, and we scan $M_1$ from $50$ GeV to $1000$ GeV.
This figure shows that independent $M_1$ can make quite a difference for a fixed $M_2$ when $M_2$ is large.
For Figure \ref{fig:amu_tanbMsoft}, a relaxation of gaugino mass unification would not make much difference since the maximum $\Delta a_\mu$ mostly happens for small $M_2$\footnote{Small $M_2$ results in a small chargino mass in Eq. (\ref{eq:charginomass}), and large $\Delta a_\mu$ in Eq. (\ref{eq:chargino_amu}).}.
In the $S$-model we do not relax $M_{1'}$ as a free parameter for the dotted curve, but rather take the NMSSM limit ($M_{1'} \gg M_2$) and relax only $M_1$.
Unless $M_{1'}$ is very small (smaller than the $M_1$ scan limit of $50$ GeV), it would not increase the maximum $\Delta a_\mu$ significantly.
For a wide range of $M_2$ (and practically for any $M_2$ in the case that $M_1$ is a free parameter), both models can produce the favored $\Delta a_\mu$.
The maximum $\Delta a_\mu$ have $\tan\beta \sim 3$, $m_{\rm smuon} \sim 100$ GeV (and $M_1 \sim 50$ GeV for the dotted curves).

\section{Conclusion}
\label{section:conclusion}
Unlike the MSSM, the electroweak symmetry breaking condition in the $S$-model requires small $\tan\beta$.
Moreover, the LEP smuon mass bound requires slepton masses above about $100$ GeV.
On the other hand, the $2.4 \sigma$ deviation of the muon anomalous magnetic moment favors large $\tan\beta$ and small slepton soft terms.
The $\Delta a_\mu$ determination gives the most severe constraint on $\tan\beta$ and slepton masses in the $S$-model; nonetheless, the $S$-model can still explain the deviation while satisfying the chargino and smuon mass limits.
Since the mass of the light smuon is constrained to be less than $180$ GeV, it would be easily observable at the next generation colliders.
It is remarkable that the parameter space that explains the $a_\mu$ deviation has a sizable overlap with the parameter space that produces an acceptable cold dark matter relic density even when only the $Z$-pole annihilation channel is considered.
The common solution space implies a lighter chargino upper mass bound of $220$ GeV, though the common solution space would increase when annihilation channels in addition to the $Z$-pole are included for the neutralino relic density calculation.
More general $U(1)'$ models that do not require small $\tan\beta$ could explain the $a_\mu$ deviation in a wider range of the parameter space.

Even though the lightest neutralino in the $S$-model is often very light, the difference of the $\Delta a_\mu$ predictions of the MSSM and the $U(1)'$ models for comparable parameters is not large.
This is because the lightest neutralino is mostly singlino-like when it is lighter than that of the MSSM, and it couples to the muon only through mixing; the properties of the other neutralinos are quite similar to those of the MSSM.
The relaxation of gaugino mass unification makes a sizable difference in both models.

The contribution of the $Z'$-loop is suppressed by the large $Z'$ mass, which is constrained by the CDF limit of $M_{Z'} \gsim 500 \sim 800$ GeV, with the limit depending on the model \cite{CDF}.
In the case that the right-handed neutrinos are $U(1)'$ charged and form Dirac particles, more severe constraints of $M_{Z'} \gsim$ multi-TeV are deduced from Big Bang Nucleosynthesis (BBN) \cite{BBN}.
Other possibilities for neutrino mass in these models are discussed in Ref. \cite{nuMass}.

\section*{Acknowledgments}

\vspace*{-1.5ex}

This research was supported in part by the U.S. Department of Energy
under Grants 
No.~DE-FG02-95ER40896,
No.~DE-FG02-04ER41305, 
No.~DE-FG02-03ER46040,
and
No.~DOE-EY-76-02-3071, 
and in part by the Wisconsin Alumni Research Foundation.


\end{document}